# NEURONAL ALIGNMENT ON ASYMMETRIC TEXTURED SURFACES

Ross Beighley[1,&], Elise M. Spedden[1,&], Koray Sekeroglu[2,&], Timothy Atherton[1], Melik C. Demirel[2,*], Cristian Staii[1,*]

1. Department of Physics and Astronomy and Center for Nanoscopic Physics, Tufts University, Medford, MA 02155
2. Materials Research Institute and Department of Engineering Science, Pennsylvania State University, University Park, PA, 16802

[*] Corresponding Authors: Prof. M. C. Demirel, E-mail: mdemirel@engr.psu.edu and Prof. C. Staii, E-mail: Cristian.Staii@tufts.edu

[&] These authors have contributed equally to the paper

**Abstract**

Axonal growth and the formation of synaptic connections are key steps in the development of the nervous system. Here we present experimental and theoretical results on axonal growth and interconnectivity in order to elucidate some of the basic rules that neuronal cells use for functional connections with one another. We demonstrate that a unidirectional nanotextured surface can bias axonal growth. We perform a systematic investigation of neuronal processes on asymmetric surfaces and quantify the role that biomechanical surface cues play in neuronal growth. These results represent an important step towards engineering directed axonal growth for neuro-regeneration studies.

Artificial growth of neurons on various substrates is of great interest for brain tissue engineering [1,2]. Neuronal cells in the brain develop two types of processes: a single, long axon that transmits information to other cells and multiple, shorter dendrites that receive electrical impulses from the axons of other neurons. Neuronal cells have been cultured on a variety of scaffolds including biopolymers, silk, and hydrogels [3] as well as grown in alignment on patterned surfaces [4,5]. These surfaces also provide physical guidance, and chemical support for neuronal cell adherence, axonal extension, network formation, and function. The axons, and in particular their dynamic unit known as the growth cone are able to detect and respond to environmental signals such as functionalization of surfaces with extracellular matrix proteins, biomolecules released by neighboring neurons at extremely low concentrations (molecular level), substrate stiffness and topographical and geometrical cues [6]. Over the past decade, there has been rapid progress in our understanding of the role played by chemical signaling and surface-based biochemical guidance on the growth cone dynamics and axonal elongation. For example, it is known that axonal navigation to their target depends on the precise arrangement of extracellular proteins on the growth surfaces [2,6,7]. It is also now recognized that mechanical interactions between neurons and their environment are playing an essential role in neuronal growth and development [5,8]. However, the neuronal response to mechanical and topographical stimuli, and the details of cell-surface interactions such as adhesion forces and traction stress generated during growth are currently poorly understood [9,10].

Directional surfaces composed of asymmetric structures are widely used in nature for wet and dry adhesion [11]. Inspired by these surfaces, Demirel *et al.* synthesized asymmetric textured surface [12] and reported an engineered nanotextured surface deriving its anisotropic adhesive wetting directly from its asymmetric nanoscale roughness [13]. In an earlier study, Demirel *et al.* studied the fibroblast adhesion and removal on directional nanofilms [14], using a fluidic shear stress to remove cells from a microfluidic channel. It has been shown that cells were removed with lower shear stresses when the flow was in the direction of nanorod tilt, compared to flow against the tilt [14]. Adhesion and retraction under asymmetric mechanical cues demonstrated unique properties [15].

Cell polarization (i.e. response to external cues such as chemical gradients and mechanical deformation) has been studied extensively on textured surfaces to understand cell fate [16]. However, unidirectional polarization in response to surface mechanical cues has not been demonstrated earlier. Here, we report axonal extension and network formation on asymmetric nanotextured surfaces. We demonstrate that axons preferentially extend along the asymmetry of the surface texture, and display an angular distribution that broadens with the increase in the surface density of the cells. We also show that a simple theoretical model, based on Brownian motion in constant field, can help to interpret the experimental results. This opens up the possibility of performing systematic studies in which the influence of different types of mechanical and topographical guidance cues could be precisely quantified. Our results could also lead to creating neural networks where signaling pathways could be studied in detail.

Figure 1 shows the schematic of growth cone adhesion and axon alignment on the directional nanotextured surface. We have been exploring nanotextured poly(chloro-p-xylylene) surfaces, which formed through vapor-phase polymerization and directed deposition of [2.2] paracyclophane derivatives as biocompatible templates. Figure 1a (top) is a schematic representation of our approach for unidirectional axon growth. Figure 1a also shows experimental results from fluorescence (middle) and bright field (bottom) imaging, which support the unidirectional model. Figure 1b shows cross sectional Scanning Electron Microscope (SEM) and Atomic Force Microscope (AFM) images of a nanotextured surface.

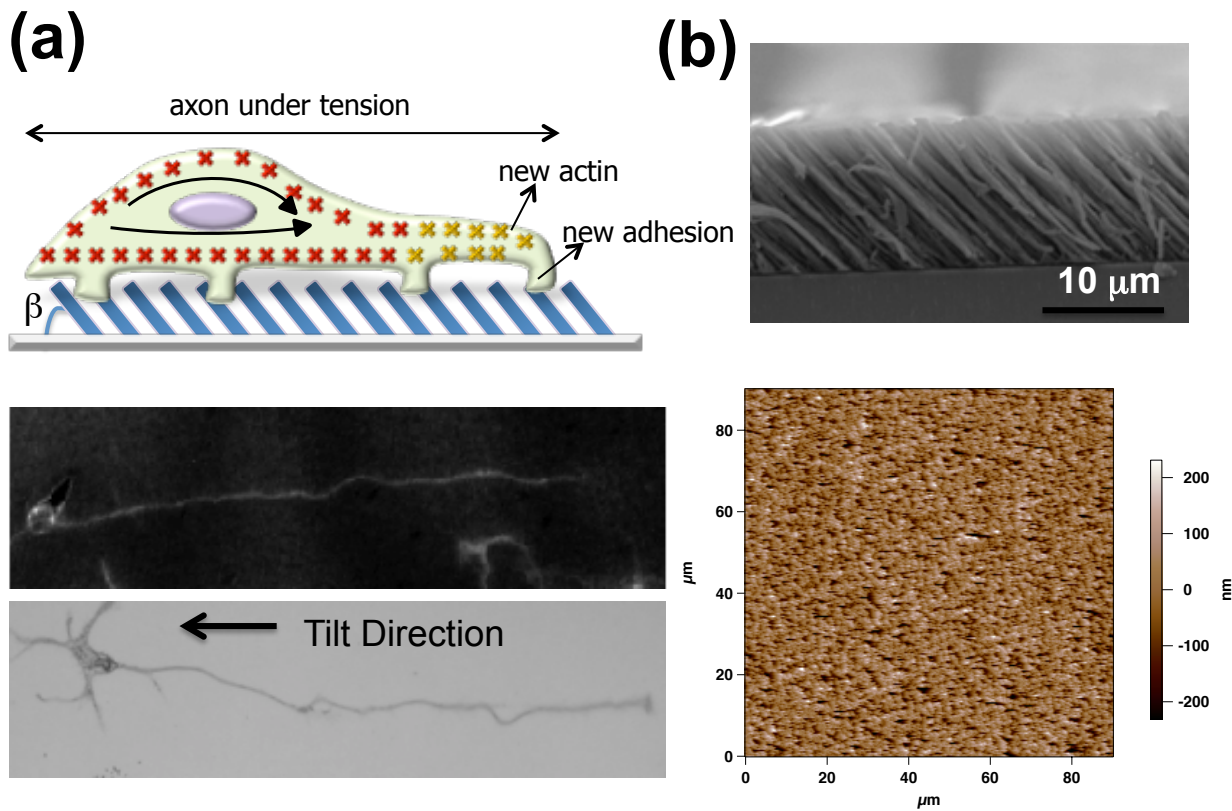

**Figure 1**. (a) *top*: Schematics of the growth cone adhesion to the asymmetric nanotextured surface, where β is the angle of the tilted nanorods. The axons extend preferentially in a direction opposite to the direction of the tilted nanorods. The direction of growth is defined as the reference direction (i.e. the 0 radians direction) for the analysis presented in the paper. Fluorescence (*middle*) and bright field (*bottom*) images of two different cells growing on the surface are also shown. (b) Scanning Electron Microscopy (*top*) and Atomic Force Microscope (*bottom*) image of the surface.

    Polymer deposition was performed using a modified Deposition System PDS 2010 (SCS, Indianapolis, IN, USA) as described earlier [12]. Glass substrates are cleaned with ethanol and acetone, and subsequently treated with a silane layer to improve the adhesion of polymer to the substrate. The film is deposited using the source material, (2,2)-dichloroparacyclophane in powder form (Uniglobe-Kisco, Whiteplains, NY, USA) to produce a directed vapor flux onto a substrate through the nozzle by tilting the substrate 10 degrees. Conventional poly(chloro-p-xylylene) deposition parameters are adopted for sublimation (175ºC) and pyrolysis (690 C and 32 Torr) of the monomer. These nanotextured surfaces were affixed to 3.5 cm glass disks using silicone glue and allowed to dry. Each surface was rinsed with sterile water, then spin-coated with 3 mL of Poly-D-lysine (PDL) (Sigma-Aldrich, St. Louis, MO) solution (0.1 mg/mL) at 1000 RPM for 10 minutes conformally. The plates were then sterilized using ultraviolet light for

≥30 minutes. The surface roughness of the nanotextured surface was measured via AFM topography before and after the application of PDL and no significant difference was observed.

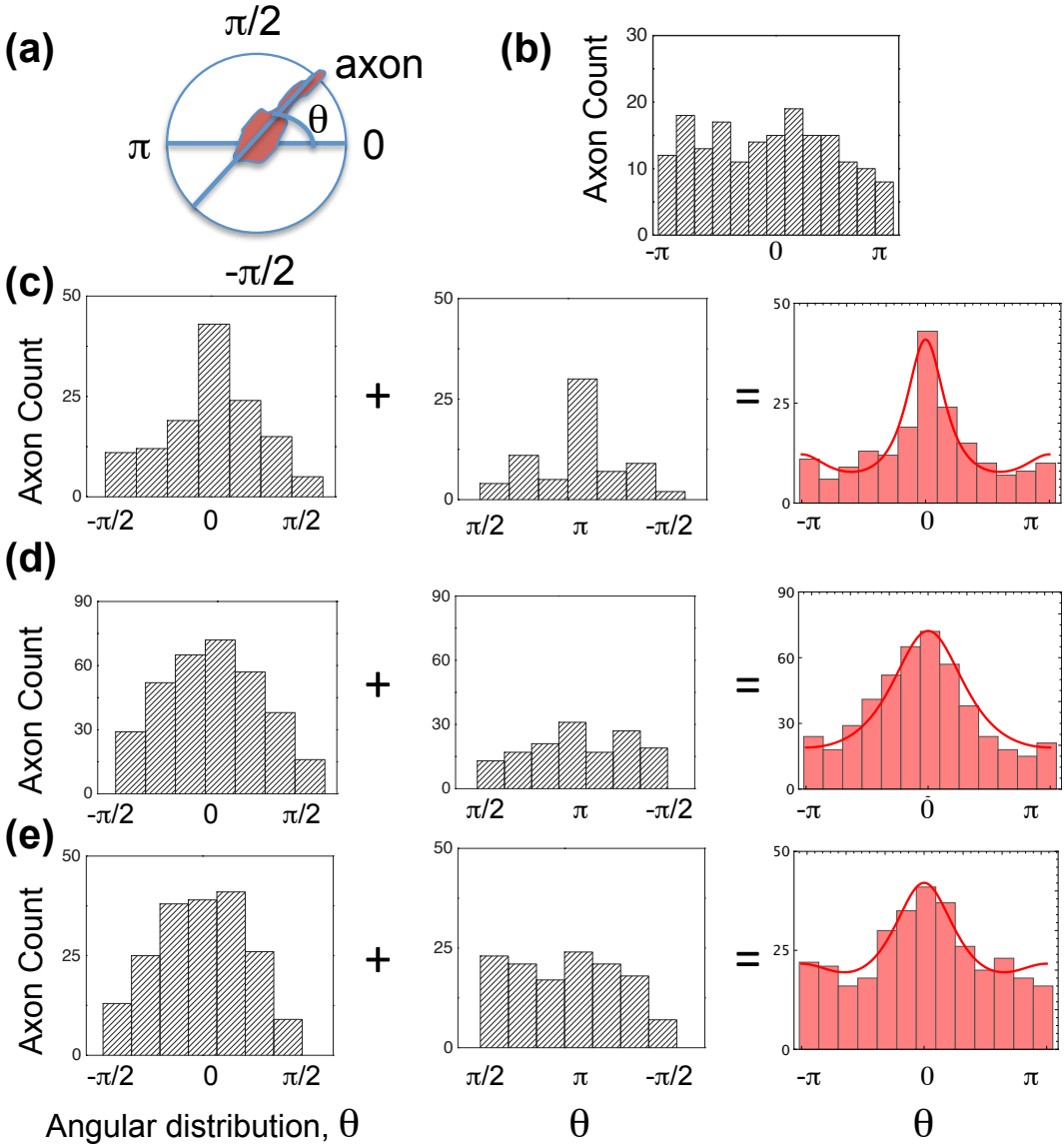

**Figure 2.** (a) The direction of axon is shown in a schematic as the direction opposite to the direction of the nanotextured surface. Angular distributions for neural growth at (b) glass surface (control) (c) 2000 cells/cm$^2$, (d) 6000 cells/cm$^2$ and (e) 25000 cells/cm$^2$. All angles are measured with respect to this direction. All plots on (c)-(e) are on the same vertical scale. The bins at π radians for each histogram shown in the middle column collect axon counts from the bins at π and -π of the corresponding histogram shown in third column. The data shows weaker preferential growth in the direction of the nanotextured surface ($\theta=\pi$) than in the opposite direction ($\theta=0$). Solid lines in the right hand column histograms represent theoretical fits using a model of Brownian motion in constant field.

Figure 2 shows the histograms of axon orientation with respect to the tilting direction of asymmetric nanorods. The bright field images are collected from a Nikon Eclipse ME 600 microscope, and the fluorescent images are collected on the inverted stage (Nikon Eclipse Ti) of an Asylum Research MFP3D Atomic Force Microscope. All images are analyzed using the Image J (http://rsweb.nih.gov/ij) computer software. The cells were grown in three different cell densities. Rat cortices were obtained from embryonic day 18 rats (Tufts Medical School). The corticies were incubated in 5 mL of trypsin at 37ºC for 20 minutes, then the trypsin was inhibited with 10 mL of neurobasal medium (Life Technologies) supplemented with GlutaMAX, b27 (Life Technologies), and antibotics (penicillin/streptomycin), containing 10 mg of soybean trypsin inhibitor (Life Technologies). The neurons were then mechanically dissociated, centrifuged, the supernatant removed, and the cells were resuspended in 20 mL of neurobasal medium containing l-glutamate (Sigma-Aldrich, St. Louis, MO). The cells were re-dispersed with a pipette, counted, and plated on the nanotextured surface at three different densities of 2000, 6000 and 25000 cells/cm$^2$. For fluorescence imaging, the live cortical samples were incubated for 30 minutes at 37ºC with 100 nM Tubulin Tracker Green (Oregon Green 488 Taxol, bis-Acetate) (Life Technologies, Grand Island, NY) in PBS. Fluorescence images were taken using a standard Fluorescein isothiocyanate (FITC) filter with excitation and emission of 495 nm and 521 nm respectively.

Figure 2 shows histograms for all experiments as well as theoretical fits from the Fokker-Plank model as explained in the next paragraph. Figure 2a show schematic of axon orientation as a function of angular distribution, $q$. Figure 2b shows angular distribution of neuronal growth on PDL/glass surface (i.e. control study). As expected the data for these samples shows non-polarized neuronal growth, with uniform angular distributions for axons. Figure 2c, 2d, and 2e show cell density histograms for 2000, 6000, and 25000 cells/cm$^2$, respectively. Axon orientation in Figure 2c shows a clear peak at 0 radians, which is defined as the direction opposite to the directionality of the nanotextured surface (see Figure 1). The data for the two low densities (2000, 6000 cells/cm$^2$, respectively) also shows a peak at $\pi$ radians, that is in the direction of the nanotextured surface. As the cell density increases, the standard deviation of the corresponding histogram increases, which reflects the fact that the axons are making more connections at higher densities, therefore deviating from the surface tilting direction. This is consistent with the fact that the neuron-neuron chemical signaling becomes more important as the cell density increases [9,17]; therefore inducing more turns in the growth cone.

To understand the axonal growth on asymmetric surfaces, we developed a simplified model based on Brownian motion in constant field. Similar models have been adopted for axon growth in the literature for symmetric surfaces [18,19]. To interpret the observed alignment of the axonal growth along the asymmetric surfaces, we derived the expected distribution of the tangent angles from a biased random walk model. The motion of the growth cone is described by the following Langevin equation, $m\dot{v} = -\alpha v + F + \xi(t)$, whre $m$ is an effective mass, t is the time, $\alpha$ is a Stokes drag coefficient, $F$ is a constant force and $\xi(t)$ is a random force which has zero mean $\langle \xi \rangle = 0$ and is Markovian $\langle \xi(t) \cdot \xi(t') \rangle = m^2 \Gamma d(t - t')$ where $\Gamma$ represents the strength of the noise and $d$ is the Dirac delta function [18]. The random force in the model causes the stochastic motion of the growth cone observed experimentally as it explores the local environment. In the absence of such a force, $\xi(t) = 0$, the model is deterministic and the equilibrium solution is motion with a constant velocity, $v_0 = F/\gamma$, where the reduced drag $\gamma = \alpha/m$ is introduced.

The directed surface enters the model in two ways: it is assumed to provide a constant force **F** and may also cause the distribution of the random forces to be uniaxial rather than azimuthally symmetric. Hence, if the alignment of the posts were parallel to the surface normal, i.e. when the angle $\beta$ between the posts and surface normal is $\beta = 0$, then $F = 0$ and $\xi(t)$ is distributed symmetrically. While it is clear on symmetry grounds that if $\beta = \pi/2$ then $F = 0$ the precise dependence of these quantities on $\beta$ requires a more detailed model of the interaction of the growth cone with the posts than pursued here.

To describe the growth anisotropy we introduce two dimensionless parameters in our model, which correspond to observables in the experiment: 1) $\kappa = v_0^2 \gamma / \Gamma$, which quantifies the strength of the effective force imparted by the surface onto the growth cone ($\kappa$ corresponds to the bias parameter in a random walk); and 2) *d*, satisfying the requirement $-1 \leq \delta \leq +1$ (see equation 1 below), which is a parameter characterizing the degree of anisotropy of the distribution of the random force field $\xi(t)$. Figure 3 provides ensembles of trajectories for various values of these two parameters.

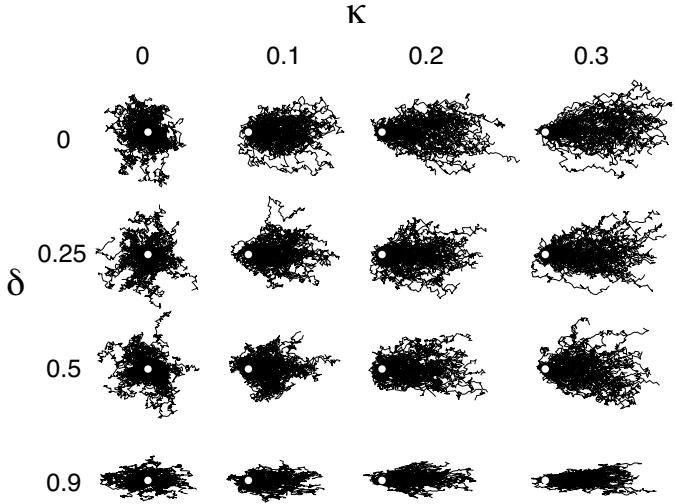

**Figure 3.** Ensembles of 100 trajectories generated from a biased, symmetric ($\delta = \kappa = 0$) and asymmetric random walk for various values of bias parameter $\kappa$ and azimuthal anisotropy parameter $\delta$. Starting points are indicated with white circles.

The motion of a dilute (non-interacting) ensemble of growth cones is described by the following Fokker-Planck equation:

$$\frac{\partial p(v,t)}{\partial t} = \nabla[\gamma(v-v_0)] \, p(v,t) + \left[(1+\delta)\frac{\partial^2}{\partial x^2} + (1-\delta)\frac{\partial^2}{\partial y^2}\right] p(v,t) \quad (1)$$

where $-1 \leq \delta \leq +1$. For *d* equals to zero, the cylindrically symmetric distribution is recovered (see Figure 3 for *d* = *k* = 0). For limiting values of *d* = ±1, the noise is purely along the x (*d* = +1) or y (*d* = -1) axes exclusively and for other values the distribution is ellipsoidal. The factors (1 + *d*) and (1 - *d*) in equation (1) are required to preserve the overall normalization of the noise. If the force **F**, and consequently $v_0$, is directed along the *x*-axis, the equilibrium solution of (1) is:

$$p(\boldsymbol{v}) = \frac{\gamma}{\pi\Gamma\sqrt{1-\delta^2}} \text{Exp}\left[-\frac{\gamma}{\Gamma}\left(\frac{(v_0-v_x)^2}{1+\delta}+\frac{v_y^2}{1-\delta}\right)\right] \quad (2)$$

Switching to polar coordinates $\boldsymbol{v} = v(\cos\theta, \sin\theta)$ and integrating over $v$, we obtain the following functional form for the distribution of the tangent angles in an ensemble of axons:

$$p(\theta) = \frac{1}{2\pi f(\theta)}\left\{e^{-\frac{\kappa}{1+\delta}}\sqrt{1-\delta^2} + \frac{\sqrt{\pi\kappa}(1-\delta)e^{-\frac{\kappa\sin^2\theta}{f(\theta)}}}{g(\theta)}\left[1 + \text{Erf}\left(\sqrt{\frac{1-\delta}{1+\delta}}\frac{\sqrt{\kappa}}{g(\theta)}\right)\right]\right\}$$
(3)

where the two functions $f(\theta) = (1 - \delta\cos 2\theta)$ and $g(\theta) = f(\theta)/\cos\theta$ have been defined together with the parameter $\kappa = v_0^2\gamma/\Gamma$.

It is instructive to examine two limiting cases for k, when $\delta = 0$. For small force where $\kappa \ll 1$ we have that:

$$p(\theta) \approx \frac{1}{2\pi} + \frac{1}{2}\sqrt{\frac{\kappa}{\pi}}\cos\theta + \cdots, \quad (4)$$

while for large force, i.e. $e^{-\kappa} \ll 1$, the distribution approaches a Gaussian:

$$p(\theta) \approx \sqrt{\frac{\kappa}{\pi}}e^{-\kappa\theta^2}. \quad (5)$$

Figure 2 shows experimental distributions obtained for various cell densities N, as well as fits of these distributions with the theoretical model (*p(q)* given by eqn. 3). Theoretical fitted values for k and d are: 0.17±0.05 and 0.44±0.08 for 2000 cells/cm$^2$, 0.15±0.03 and 0.06±0.06 for 6000 cells/cm$^2$, and 0.04±0.01 and 0.17±0.03 for 25000 cells/cm$^2$ respectively. The fitted value of k clearly decreases proportionally with increasing cell seeding density N. This is expected as k represents a balance between the applied force on the growth cones, which is a property of the surface and hence should be identical for each experiment, and the random forces from chemotactic signaling described by the reduced drag term g. These fitted values are consistent with the physical expectation that g ~ N. The asymmetry parameter d was found to vary between 0.06 and 0.44, which indeed confirms that the presence of the surface induces an anisotropy in the chemotactic signaling. We note that the asymmetry parameter, d, should be only considered as a theoretical qualitative estimate, which is needed in order to introduce anisotropy in the distribution of the random force field. However, our simple model captures the main features of axonal elongation on these asymmetric surfaces.

In conclusion, we have demonstrated that a unidirectional surface can bias axonal growth. By varying the density of neuron cells on asymmetric textured surfaces, we showed the competition between mechanical and chemical (neuronal signaling) on these surfaces. The unidirectional mechanical cues dominate cell growth at low cell densities, while the spatially symmetric chemical cues start to play an increasingly important role at higher densities. We also note that directional axonal growth with mechanical cues has potential applications in peripheral and spinal cord injuries. We plan to extend our studies to high-throughput assays for neural cue studies in the near future.

**Author Contributions:** MCD and CS planned and supervised the research. KS prepared the nanotextured surfaces, EMS and RB studied the neural growth, TA developed the theoretical model for axonal growth. All authors contributed to writing and revising the manuscript, and agreed on its final contents.